Correction: The reference [6] is Andrew's convex hull algorithm rather than the Melkman algorithm.

# A Straightforward Preprocessing Approach for Accelerating Convex Hull Computations on the GPU

Gang Mei

Institute of Earth and Environmental Science, University of Freiburg

Albertstr.23B, D-79104, Freiburg im Breisgau, Germany

gang.mei@geologie.uni-freiburg.de    gangmeiphd@gmail.com

**Abstract** An effective strategy for accelerating the calculation of convex hulls for point sets is to filter the input points by discarding interior points. In this paper, we present such a straightforward and efficient preprocessing approach by exploiting the GPU. The basic idea behind our approach is to discard the points that locate inside a convex polygon formed by 16 extreme points. Due to the fact that the extreme points of a point set do not alter when all points are rotated in the same angle, four groups of extreme points with min or max $x$ or $y$ coordinates can be found in the original point set and three rotated point sets. These 16 extreme points are then used to form a convex polygon. We check all input points and discard the points that locate inside the convex polygon. We use the remaining points to calculate the expected convex hull. Experimental results show that: when employing the proposed preprocessing algorithm, it achieves the speedups of about 4x ~5x on average and 5x ~ 6x in the best cases over the cases where the proposed approach is not used. In addition, more than 99% input points can be discarded in most experimental tests.

## 1. Introduction

The finding of convex hulls is a fundamental issue in computer science, which has been intensively studied for many years. When calculating the convex hull for a large set of points, an effective strategy for improving the computational efficiency is to discard the interior points that have been exactly determined. This strategy is referred to as the *preprocessing* procedure. The most commonly used preprocessing approach is to form a convex polygon or polyhedron using several determined extreme points first and then discard those points that locate inside the convex polygon or polyhedron; see [1, 2]. The simplest case is to form a convex quadrilateral using four extreme points with min or max $x$ or $y$ coordinates and then to check each point to determine whether it locates inside the quadrilateral; see [3]. Recently, Cadenas and Megson [4] presents a linear and general preprocessing approach which does not require an explicit sort of points.

In this paper, an efficient preprocessing approach is proposed, which is well suitable for being implemented on the GPU. This approach is similar to the simple preprocessing method introduced in [3]. The basic idea behind this approach is also to discard those interior points that locate inside a convex polygon formed by extreme points. However, in the proposed method typically 16 rather than 4 extreme points are used to form the convex polygon.

## 2. Methods

The basic ideas behind our algorithm are quite simple. The first idea is to filter the points by discarding those interior points that locate inside a convex polygon formed by several extreme points. The second idea is that we can first rotate a set of points and then find those extreme points with the min or max $x$ or $y$ coordinates. Those points having the min or max $x$ or $y$ coordinates are obviously extreme ones. It is clear that the extreme points of a point set do not alter when all points in the set are completely rotated the same angles. Therefore, it is easily able to obtain three groups of extreme points by rotating a set of points in 30, 45, and 45 degrees; see Figure 1. Together with the group of extreme points of the original point set, there are four groups of extreme points and in total 16 (4 * 4) extreme points. These extreme points can be used to form a convex polygon; and then each of the rest points is checked to determine whether it falls in the convex polygon. Those points locating inside the convex polygon are definitely interior points, and can be discarded directly.

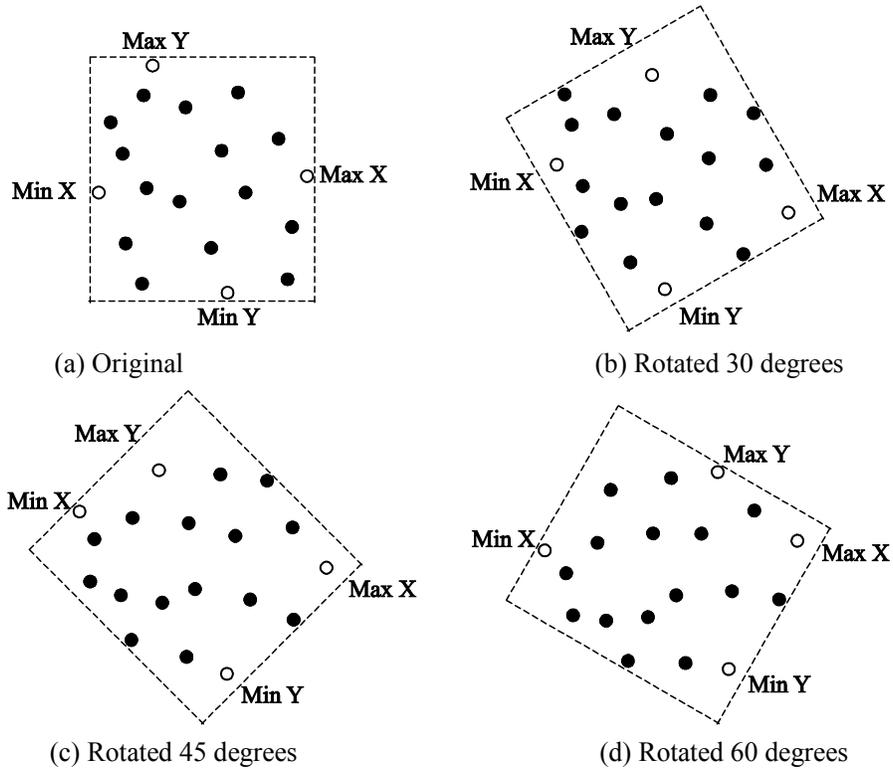

(a) Original  (b) Rotated 30 degrees

(c) Rotated 45 degrees  (d) Rotated 60 degrees

**Figure 1** Original and rotated point sets and extreme points with min or max coordinates

The procedure of our algorithm is listed as follows:

Step 1: Locate extreme points

The simplest approach for locating the extreme points for a set of points is to find those points with min or max *x* or *y* coordinates. We first find typically four extreme points of the original point set. We then rotate the set of points in 30, 45, and 45 degrees and find a group of typically four extreme points for each version of rotated point set. In general, there are totally 16 extreme points that can be found. In some cases, the number of extreme points is less than 16 due to duplicate points. Several simple CUDA kernels are designed to perform the rotating of point sets; and the finding of extreme points is realized by using the efficient parallel reduction primitive provided by *Thrust* [5].

Step 2: Form a convex polygon

We use the Andrew's algorithm [6] to form the convex hull of the 16 extreme points. This part of the work is performed on the CPU.

Step 3: Discard interior points

Those points locating inside the convex polygon formed by the extreme points need to be discarded. We design a CUDA kernel to carry out the discarding in parallel. Each thread is responsible for checking a point to determine whether it falls in the convex polygon. To speed up the calculation, the coordinates of all extreme points are first transformed from global memory to shared memory and then accessed by all the threads within the same thread block.

## 3. Results

To evaluate the effectiveness of our algorithm, we compare the running time of finding convex hulls in two cases: (1) the proposed preprocessing algorithm CudaPre is not adopted, and the original point set is directly used to calculate the convex hull; (2) the algorithm CudaPre is first employed to filter the original set of points, and then the remaining points are used to find the convex hull.

We have tested our algorithm by employing the Qhull library [7] on the following platform. The adopted machine features an Intel i5-3470 processor (3.20GHz), 8GB of memory and a NVIDIA GeForce GT640 (GDDR5) graphics card. The graphics card GT640 has 1GB of RAM and 384 cores. We have used the CUDA toolkit version 6.0 on Window 7 Professional to evaluate all the experimental tests.

We have created three groups of datasets for testing. The first group includes 5 sets of randomly distributed points in a square that are generated using the `rbox` component in Qhull. Similarly, the second group is composed of 5 sets of randomly distributed points in a circle. The third group consists of 5 point sets that are derived from 3D mesh models by projecting the vertices of each 3D model onto the XY plane. These mesh models listed in Table 3 are directly obtained from the Stanford 3D Scanning Repository[1] and the GIT Large Geometry Models Archive[2].

The running time of the above three groups of test data is listed in Tables 1 ~ 3. The speedups in the case where CudaPre is adopted over the case where CudaPre is not employed is about 4x ~ 5x on average, and 5x ~ 6x in the best cases. For the sets of points locating in squares, the algorithm CudaPre achieves the best results; see Table 1. In addition, the effectiveness of discarding interior points using CudaPre can be obviously observed: for the first and the third groups of test data, more than 99% input points are discarded, while more than 96% points are removed from the second group of test data (Table 2).

Table 1 Comparison of running time (/ms) for the point sets locating in squares

| Size | Qhull | Qhull + CudaPre | | | Remaining Points (%) | Speedup |
|------|-------|------|--------|-------|------|------|
|      |       | Total | CudaPre | Qhull | | |
| 1M   | 106   | 27.2  | 26.2    | 1     | 0.06 | 3.90 |
| 2M   | 203   | 44.4  | 43.4    | 1     | 0.06 | 4.57 |
| 5M   | 515   | 90.7  | 89.7    | 1     | 0.06 | 5.68 |
| 10M  | 1032  | 163.4 | 162.4   | 1     | 0.03 | 6.32 |
| 20M  | 2216  | 323.9 | 322.9   | 1     | 0.02 | 6.84 |

Table 2 Comparison of running time (/ms) for the point sets locating in circles

---

[1] http://www-graphics.stanford.edu/data/3Dscanrep/

[2] http://www.cc.gatech.edu/projects/large_models/

Correction: The reference [6] is Andrew's convex hull algorithm rather than the Melkman algorithm.

| Size | Qhull | Qhull + CudaPre | | | Remaining Points (%) | Speedup |
|---|---|---|---|---|---|---|
| | | Total | CudaPre | Qhull | | |
| 1M | 132 | 42.8 | 26.8 | 16 | 3.46 | 3.08 |
| 2M | 258 | 61.6 | 45.6 | 16 | 3.44 | 4.19 |
| 5M | 653 | 131.9 | 96.9 | 35 | 3.39 | 4.95 |
| 10M | 1335 | 253.9 | 180.9 | 73 | 3.39 | 5.26 |
| 20M | 2661 | 505.7 | 352.7 | 153 | 3.39 | 5.26 |

**Table 3** Comparison of running time (/ms) for the point sets derived from 3D mesh models

| Model | Size | Qhull | Qhull + CudaPre | | | Remaining Points (%) | Speedup |
|---|---|---|---|---|---|---|---|
| | | | Total | CudaPre | Qhull | | |
| Blade | 0.8M | 83 | 25.8 | 24.8 | 1 | 0.52 | 3.22 |
| Vellum | 2.1M | 215 | 46.9 | 45.9 | 1 | 0.14 | 4.58 |
| Asian Dragon | 3.6M | 344 | 76.4 | 72.4 | 4 | 0.74 | 4.50 |
| Thai Statue | 5M | 468 | 94.1 | 92.1 | 2 | 0.12 | 4.97 |
| Lucy | 14M | 1304 | 252.6 | 247.6 | 5 | 0.25 | 5.16 |

## 4. Discussion

### 4.1. Effectiveness of Filtering

The experimental results presented in Tables 1 ~ 3 show that the effectiveness of discarding interior points using CudaPre for three groups of test data are different, especially for the first and the second groups. After discarding interior points using CudaPre, much less remaining points still exist for the first group than those for the second groups. These results are probably caused by the following facts.

The convex hull of the points locating in a square is typically an approximate square; and the convex hull of the points locating in a circle is in general a polygon which is like a circle. When applying CudaPre, a convex polygon needs to be formed using 16 extreme points. The convex polygon is quite close to the expected convex hull, i.e., the approximate square, for the points locating in a square. Thus, nearly all of the input points fall in the convex polygon formed by the extreme points and would be discarded. Very few points remain in this case. For the points locating in a circle, a convex polygon composed of 16 extreme points is also close to the expected convex hull (i.e., a polygon like a circle); but the convex polygon is not fine enough to represent the polygon that is like a circle. The region bounded by the expected convex hull is larger than that covered by the convex polygon. Thus, there are still many points that locate outside the convex polygon formed by 16 extreme points.

For the general cases where points are not distributed in a regular square or circle, the effectiveness in the use of CudaPre is worse than that in the best cases (points locating in a square), but much better than that in the worst cases (points locating in a circle). These general cases can be considered as the transitional phase between the worst cases and the best cases.

### 4.2 Data Dependency

In the Step 1 and Step 3 of CudaPre, there are no data dependency issues. Thus, the above two steps can be well mapped to the massively parallel nature of the modern GPU. For example, the finding of points with min or max coordinates can be realized using the efficient parallel reduction provided by Thrust [5]; and the rotating of points can be performed by invoking a CUDA kernel where each thread is responsible for calculating the new coordinates for one point; in another kernel, each thread takes the responsibilities for checking a point to determine whether it locates inside the convex polygon. In CudaPre, the only step that has data dependency issues is the forming of a convex polygon using 16 extreme points. Fortunately, this step is simple and easy to implement on the CPU. For the entire algorithm, the feature of having less data dependencies makes it simple and easy to implement in practical applications.

### 4.3 Complexity

The time complexity of CudaPre is $O(n)$. The finding of extreme points with min or max coordinates, the rotating of a set of points, and the determining of interior points completely run in $O(n)$. And the forming of a convex polygon using 16 extreme points needs constant time. Thus, the entire algorithm runs in $O(n)$ time.

## 5. Conclusion and Outlook

We have presented a straightforward preprocessing method for accelerating the finding of convex hulls for planar point sets. The basic idea behind our algorithm is to discard the interior points that locate inside a convex polygon formed by 16 extreme points. These extreme points with min or max coordinates are found by rotating the input point set in 30, 45, and 60 degrees. We have evaluated our algorithm, CudaPre, by comparing the efficiency when employing the proposed algorithm with the efficiency in the case where our algorithm was not adopted. Our results indicate that: the speedups in the case where CudaPre is adopted over the case where CudaPre is not employed is about 4x ~5x on average, and 5x ~ 6x in the best cases. In addition, when using CudaPre, more than 99% input points can be discarded in most tests.

The proposed algorithm is only applicable in 2D. However, it can be very easily extended to the three dimensions. In 3D, typically six extreme points can be obtained by finding those points with the min or max $x$, $y$, or $z$ coordinates. More groups of six extreme points can also be found after rotating the set of points along a specific axis. These extreme points can be then used to form a convex polyhedron. Those points locating inside the convex polyhedron must be interior points, and can be directly discarded.

Correction: The reference [6] is Andrew's convex hull algorithm rather than the Melkman algorithm.